\RequirePackage[2020-02-02]{latexrelease}
\documentclass[superscriptaddress,secnumarabic,
amssymb,amsmath,nobibnotes,aps,prd,showkeys,showpacs,nofootinbib]{revtex4}%
\usepackage{graphicx}
\usepackage{subfigure}
\usepackage{subcaption}
\usepackage{wrapfig}
\usepackage{lipsum}
\usepackage{epstopdf}
\usepackage{epsf}
\usepackage{epsfig}
\usepackage{scrtime}
\usepackage{multirow}
\usepackage[multiple]{footmisc}
\usepackage{amsfonts, graphics}
\usepackage{bm}
\usepackage[utf8x]{inputenc}
\usepackage{amsmath}
\usepackage{amsfonts}
\usepackage{amssymb}
\usepackage{color}
\usepackage{natbib}%
\providecommand{\U}[1]{\protect\rule{.1in}{.1in}}
\newcommand{\be}{\begin{equation}}
\newcommand{\ee}{\end{equation}}

\newcommand{\mincir}{\raise
-3.truept\hbox{\rlap{\hbox{$\sim$}}\raise4.truept\hbox{$<$}\ }}
\newcommand{\magcir}{\raise
-3.truept\hbox{\rlap{\hbox{$\sim$}}\raise4.truept\hbox{$>$}\ }}

\begin{document}

\title{Dynamics of Dissipative Gravitational Collapse in the Morris-Thorne Wormhole Metric: One Scenario - Several Outcomes }
\author{Subhasis Nalui}
\email{subhasis222nalui@gmail.com }

\author{Subhra Bhattacharya}
\email{subhra.maths@presiuniv.ac.in}
\affiliation{Department of Mathematics, Presidency University, Kolkata-700073, India}

\keywords{Morris-Thorne Metric, Heat quotient, Collapse, Junction condition, Vaidya Metric}
\pacs{04.20.cv, 98.80.-k.}

\begin{abstract}

We consider the dynamical Morris-Thorne metric with radiating heat flow. By matching the interior Morris-Thorne metric with an exterior Vaidya metric we trace out the collapse solutions for the corresponding spherically symmetric inhomogeneous distribution of matter. The solutions obtained are broadly of four different types, giving different end state dynamics. Corresponding to three of the solutions we elaborate the collapsing dynamics of the Morris-Thorne type evolving wormhole. We show that for all those cases where collapse upto zero proper volume is obtained in finite time, the ensuing singularity is always a black hole type. However our solutions can also show other end states, like oscillating wormhole-black hole pair or infinite time contracting universe or a conformal past matter dominated universe. In all the cases we have worked out the background dynamics and physics of the solution. All our solutions are illustrated with appropriate graphical descriptions.
\end{abstract}
\maketitle

\section{Introduction}

Gravitational collapse has been dealt in the literature quite extensively. Yet contemporary research on general relativity (GR) sees no dearth of interest on the topic. This is due to the plethora of possibilities and questions that it can resolve and as well raise. Interests in gravitational collapse could be traced back to as early as 1930's when the upper limit of stable stellar mass was established by the Chandrashekhar limit \cite{chand}. The natural question then was the fate of stellar objects with masses greater than Chandrashekhar limit. Soon research revealed the mathematics of gravitational collapse in the works of Oppenheimer, Snyder \cite{os} and Datt \cite{dat} (also called OSD models). The topic received an impetus since 1970's, when space time singularities were found as a generic feature of GR. It was during this period the actual mathematics of stellar gravitational collapse was worked out. The outcome of stellar gravitational collapse was inevitably either the black hole or the naked singularity \cite{yod, sze, ear, or, jos1, jos2}. In the past decade black holes have been observationally verified in the universe. But existence of naked singularity is controversial and as yet speculative physics. Gravitational collapse has been treated from a number of perspective, like varying space-time backgrounds, in scalar field theory \cite{christ}, in electromagnetic or magnetic field \cite{thorn1, pris}, in different gravity theories like $f(R)$ gravity \cite{jar, gos}, Lovelock gravity \cite{oha}, Horava-Lifshitz gravity \cite{green}, Gauss-Bonnet gravity \cite{ban1}, in the presence of inhomogeneous background metric \cite{jos3, mena1, lasky, ud, sc}, anisotropies and radiating heat loss \cite{herr2, herr3, herr4, shar1, shar2} or in case of radiating luminosity \cite{sch} etc.

In this article we have considered a scenario of dissipative gravitational collapse. Dissipative processes in collapse is an established phenomenon. It was argued in \cite{am} that Newtonian forces result in thermal radiations of a collapsing gravitating ball of mass. The importance of dissipative collapse can be understood from the number of different scenarios where it was examined. Vaidya \cite{vad} in his pioneering work considered gravitational collapse in the presence of outgoing thermal radiations and the outcome was the Vaidya metric. The Vaidya metric has been studied in a variety of scenario. In most of the literature on collapse an exterior Vaidya metric is matched with a suitable interior metric by fixing the junction conditions \cite{san}. Here we will consider a case of dissipative collapse, where the exterior solution supporting thermal radiations will be the Vaidya metric. 

The matter stress tensors is additionally assumed as inhomogeneous and anisotropic. Here we are investigating the dynamic scenario of collapse with dissipation, which naturally contemplates directionally varying principal stresses. Anisotropy in pressure is likely to introduce instability in the system triggering a plausible collapse or expansion. Indeed it was shown in \cite{herr5} that in an initially shear free system presence of dissipation can induce anisotropy in an isotropic system. There are several physical motivations in assuming anisotropic principal stresses. In the past anisotropic pressures have been associated with varying surface red-shift and critical mass of compact objects. Ultra dense Neutron and Boson stars have been found stable under pressure anisotropy when the tangential stress exceeds the radial stresses \cite{kd}. Magnetic fields like the Fermi gas fields and viscosity in fluid are examples of naturally anisotropic matter. Idea of anisotropic pressure is now more than a century old and it was Jeans who put forth the concept in the Newtonian setting \cite{jean}. Later historical developments were due to the work of Lemaitre \cite{lem} followed by Bowers and Liang \cite{bl}. A complete review of the developments in anisotropic matter could be found in  \cite{herr1} and \cite{jp}.

Some time back anisotropy was associated with the evolution of structure scalars and Weyl tensors \cite{herr6}. Very recently the relation between Weyl tensors and anisotropy facilitated the development of a new classification of compact astrophysical objects based on the ``evolution of fluid distribution" called the complexity factor \cite{herr7}. In a previous work \cite{herr8} it was shown that in homologous conditions a shear free, non-dissipative flow will be geodesic and that the complexity for such flows is always zero with the ensuing system being stable undergoing homogeneous expansion/contraction. Whereas no such conclusion could be drawn for dissipative, non-geodesic flows, even under the vanishing complexity case. This is relevant in our context for we shall study our system in a inhomogeneous, anisotropic, dissipative but shear free framework where stable systems are associated with zero complexity and geodesic flows.

We will consider the dissipative, dynamical gravitational collapse in the inhomogeneous Morris-Thorne (MT) space time. The inhomogeneous MT space-time was used by Morris and Thorne \cite{mt} to describe the hypothetical geometry of wormholes (sometimes the MT space-time has also been named the IFRW or the inhomogeneous FRW metric \cite{sc}, because MT metric essentially describes an inhomogeneous spherically symmetric metric which is asymptotically flat \cite{sb1}). In past, gravitational collapse has been dealt in the inhomogeneous LTB metric and in the Szekeres metric \cite{ud}. The MT metric has been used extensively in the literature from a variety of different perspectives. In most cases it was used to describe the traversable wormhole \cite{mtw}. A traversable wormhole is characterized by a tube like structure called the throat that has matter violating the null energy conditions (NEC). Which means that at the wormhole throat one gets defocussing null geodesics. At the time of its inception, authors in \cite{mt} used this geometric manipulation on the metric to make it amenable for traversal.  However outpouring number of research focussed on this ``bad feature" of the MT wormhole and aimed to do away with this. There are now several examples of the dynamic wormhole metric existing with matter satisfying NEC \cite{wnec}. Then there are static wormholes in higher gravity theories that can exist in normal matter \cite{sb2,sb3,sb4}. Recent work on wormholes suggest its possible detection mechanism \cite{wdetc}. 

Recently a modified representation of the MT metric and Schwarzschild metric \cite{sv} was used in \cite{sksc} to show that gravitational collapse can lead to formation of a static wormhole configuration. \cite{sus} also considered a collapsing wormhole geometry. Here they used Tolman's collapsing dust model for the wormhole geometry to show gravitational collapse. In \cite{sc} the dynamic MT metric was used to depict gravitational collapse without invoking the throat exoticity or dissipation. In this article we consider the dynamic MT wormhole metric with dissipation. There are various descriptions of the dynamic wormhole throat using geometric manipulations, like defining the throat as the temporal outer trapping horizon or anti-trapped surfaces \cite{vis, hay} with NEC violation as a generic property while \cite{mae} stated NEC violation could be avoided by describing the geometry as a trapped space like hypersurface with no trapping horizon. Given the variety of possibilities on the dynamic wormhole geometry we analyse the collapse dynamics under the influence of dissipative radial heat flow.    

The paper is organised as follows: In section 2 we briefly describe the evolving MT wormhole metric, which is our interior configuration. In section 3 we describe the exterior Vaidya metric. The corresponding matching of the junction conditions at the interface of the two metric is developed in Section 4. In section 5 we give the details of solving the dynamical equation and in section 6 the corresponding solutions. In subsections 6.1-6.4 we discuss in details the mathematical and physical interpretations of the solutions obtained. We also present graphical representation of all the solutions here. Finally we present a brief conclusion in section 7.

\section{The Morris-Thorne Metric: Interior Space-Time}

We consider spherically symmetric distribution of the radiating fluid described by the the MT metric 
\begin{equation}
dS_{-}^{2}=-e^{2\phi(r,t)}dt^{2}+a^{2}(t)\left(\frac{dr^{2}}{1-b(r)/r}+r^{2}(d\theta^{2}+\sin^{2}\theta d\psi^{2})\right)\label{intmetric}
\end{equation}
where the coordinates are identified as $x^{0}=t,x^{1}=r,x^{2}=\theta$ and $x^{3}=\psi.$ The radial coordinate $r$ is not the radial distance, instead it indicates the radius of a sphere located at the wormhole throat. In the static wormhole, the throat being identified at some $r=r_{0}$ such that the function $b(r),$ called the throat function or the shape function is $b(r_{0})=r_{0}.$ The same analysis can be carried forward in the evolving wormhole scenario. The throat can be identified by defining the Euclidean embedding of an equatorial slice of the above metric corresponding to some fixed time $t.$ For the above case, on the embedding the surface of the two sphere will be given by $\mathcal{R}(t,r)=a(t)r.$ Comparing the above metric at the equatorial slice for some fixed time $t$ with the Euclidean embedding, we get 
\begin{align*}
\frac{d\mathcal{R}}{dz}&=\pm\left(\frac{r}{b(r)}-1\right)^{\frac{1}{2}}\\
\frac{d^{2}\mathcal{R}}{dz^{2}}&=\frac{1}{2a(t)}\left(\frac{b(r)-rb'(r)}{b^{2}(r)}\right).
\end{align*}
Now using the definition of throat as the point where $\frac{dz}{d\mathcal{R}}$ diverges we get $b(r)=r.$ This also corresponds to the minimum value of $\mathcal{R}(z)$ on the embedding as seen by $\frac{d\mathcal{R}}{dz}=0$ and $\frac{d^{2}\mathcal{R}}{dz^{2}}>0$ at some radius $r=r_{0}$ for which $b(r)=r.$ As a result of this the radial coordinate ranges between $(r_{0},r).$ The function $\phi(r,t)$ is the redshift function. The metric (\ref{intmetric}) will represent a traversable wormhole provided it does not have a horizon. This means that the function $e^{\phi(r,t)}$ must be regular for all $r$ and at all time $t.$ Since we are considering a dissipative heat flow, the corresponding energy momentum tensor $T_{\mu\nu}$ is defined as:
\begin{equation}
T_{\mu\nu}=(\rho+p_{t})u_{\mu}u_{\nu}+p_{t}g_{\mu\nu}+(p_{r}-p_{t})\chi_{\mu}\chi_{\nu}+q_{\mu}u_{\nu}+q_{\nu}u_{\mu}
\end{equation}
where the $\rho,p_{r},p_{t}$ and $q$ are the energy density, radial pressure, tangential pressure and the radial heat flux respectively. $u_{\mu}$ is the unit normal four velocity, while $\chi_{\mu}$ is the space like unit vector. Heat flux $q,$ the velocity vector $u^{\mu}$ and the space like vector $\chi_{\mu}$ are related by $u^{\mu}u_{\mu}=-1,~q_{\mu}u^{\mu}=0,~\chi^{\mu}\chi_{\mu}=1$ and $\chi^{\mu}u_{\mu}=0.$ Assuming this to be the interior space-time configuration, the Einstein's field equations are given by:
\begin{align}
\kappa \rho(r,t)=&3H^{2}e^{-2\phi}+\frac{b'(r)}{a^{2}r^{2}}\label{fe1}\\
 \kappa p_{r}(r,t)=&-(2\dot{H}+3H^{2}-2H\phi_{t})e^{-2\phi}+2\left(1-\frac{b}{r}\right)\frac{\phi_{r}}{a^{2}r}-\frac{b}{a^{2}r^{3}}\label{fe2}\\
 \kappa p_{t}(r,t)=&-(2\dot{H}+3H^{2}-2H\phi_{t})e^{-2\phi}+\left(1-\frac{b}{r}\right)\left[\frac{\phi_{rr}+\phi_{r}^{2}}{a^{2}}-\frac{b'r+b-2r}{2a^{2}r(r-b)}\phi_{r}-\frac{b'r-b}{2a^{2}r^{2}(r-b)}\right]\label{fe3}\\
\kappa q(r,t)=&-2He^{-\phi}\phi_{r}\frac{r-b}{a^{2}r} \label{fe4}
\end{align}
where $a(t)$ is the scale factor with $H=\frac{\dot{a}}{a}$ being the Hubble parameter. $Dot$ and $prime$ representing differentiation with respect to (w.r.t) $t$ and $r$ respectively (note that $\phi_{t}$ is partial differentiation w.r.t $t$ and $\phi_{r}$ partial differentiation w.r.t $r$). Here $\kappa$ is the dimensionless gravitational coupling constant (for all subsequent analysis we take $\kappa=1.$) The peculiarity of the above MT space time metric is that it usually defines a NEC violating distribution of matter. Our aim is to study the collapsing solutions, if any, corresponding to such a distribution of matter. 

It may be noted that recently the above wormhole metric, for the static case, has been classified as ``complex" \cite{sn} based on the new complexity factor classification suggested in \cite{herr7}. For the most general class of the above metric with $\phi_{r}\neq 0$ homologous distribution does not give geodesic flow and hence the complexity factor $y_{tf}\neq 0.$ For the above dynamic metric the complexity factor is given as: $y_{tf}=\left(1-\frac{b}{r}\right)\frac{\phi_{rr}+\phi_{r}^{2}-r^{-1}\phi_{r}}{a^{2}}-\frac{b'r-b}{2a^{2}r^{2}}\phi_{r}.$ Clearly $y_{tf}$ vanishes only when $\phi_{r}=0$ that is, for homologous, shear free, non-dissipative fluid (follow (\ref{fe4})) which gives geodesic flow. Thus by \cite{herr8} the above considered inhomogeneous, anisotropic and non-dissipative wormhole geometry is stable only if $\phi_{r}=0\Leftrightarrow y_{tf}=0.$ However for homologous condition the above metric, with $\phi_{r}\neq 0,$ will never give geodesic flow, hence with or without dissipation no conclusive statement can be made about its stability \cite{herr7,herr8}. 

\section{The Vaidya Metric: Exterior Space-Time}

To order to study the collapsing solutions for a heat dissipating distribution of matter, we match it with an exterior Viadya metric given by:
\begin{equation}
dS_{+}^{2}=-\left(1-\frac{2M(v)}{R}\right)dv^{2}-2dvdR+R^{2}(d\theta^{2}+\sin^{2}\theta d\psi^{2})\label{extmetric}
\end{equation}
The coordinate $x^{0}=v$ is identified as the retarded or outgoing null metric, with $M(v)$ a function associated with it, also identified as the mass function of the enclosed gravitating body. The remaining coordinates are $x^{1}=R,x^{2}=\theta$ and $x^{3}=\psi.$

\section{The Junction Conditions}

We match the interior MT space time (\ref{intmetric}) with the exterior Vaidya metric (\ref{extmetric}) at the junction interface identified by the hypersurface $\Sigma.$ Then on this $\Sigma$ hypersurface one must satisfy the smooth matching of the exterior and interior metric as well as the smooth matching of the extrinsic curvature tensors, which is given by:
{\setlength{\mathindent}{0cm}
\begin{equation}
\hspace{-20em}\text{First Funfdamental Form:}~~dS^{2}_{-}\mid_{\Sigma}=dS^{2}_{+}\mid_{\Sigma}=dS^{2}_{\Sigma}
\end{equation}
\begin{equation}
\hspace{-25em}\text{Second Fundamental Form:}~~K^{-}_{ij}=K^{+}_{ij}
\end{equation}
where $dS^{2}_{\Sigma}$ is the metric defining the surface $\Sigma$ and $K_{ij}^{\pm}$ are the corresponding extrinsic curvatures on the interior and exterior metric and defined as $K_{ij}^{\pm}=-n^{\pm}_{\alpha}\frac{\partial^{2} x^{\alpha}}{\partial\xi^{i}\partial\xi^{j}}-n^{\pm}_{\alpha}\Gamma^{\alpha}_{\gamma\beta}\frac{\partial x^{\gamma}}{\partial\xi^{i}}\frac{\partial x^{\beta}}{\partial\xi^{j}}.$ Here $n^{\pm}_{\alpha}$ are the unit normal vectors to the surface $\Sigma,~\Gamma^{\alpha}_{\gamma\beta}$ are the Christoffel symbols, the coordinates $\xi^{i}$ are the corresponding coordinates on $\Sigma$ and defined by $\xi^{i}=(\tau, \theta,\psi)$ while $x^{\alpha}$ the coordinates on the exterior and interior metric, defined above. The metric on the hypersurface is given by:
\begin{equation}
dS^{2}_{\Sigma}=-d\tau^{2}+\mathsf{R}^{2}(\tau)(d\theta^{2}+\sin^{2}\theta d\psi^{2})\label{sig}
\end{equation} 
such that the hypersurface in $(t,r,\theta,\psi)$ coordinates is defined at the constant radius $r=r_{b}.$ Now comparing (\ref{intmetric}) and (\ref{extmetric}) with (\ref{sig}) we get the following relations on the hypersurface $\Sigma$
\begin{align}
\frac{d\tau}{dt}&=e^{\phi(r,t)}\label{bc1}\\
ra(t)&=\mathsf{R}(\tau)=R(v)\label{bc2}\\
\left(\frac{d\tau}{dv}\right)^2&=1-\frac{2M}{R}+2\frac{dR}{dv}\label{bc3}
\end{align}
The normal vector to the interior and exterior of the surface $\Sigma$ is given by
\begin{align}
n^{-}_{\alpha}&=\left(0,\frac{a(t)}{\sqrt{1-\frac{b(r)}{r}}},0,0\right)\\
n^{+}_{\alpha}&=\left(1-\frac{2M}{R}+2\frac{dR}{dv}\right)^{-\frac{1}{2}}\left(-\frac{dR}{dv},1,0,0\right)
\end{align}
while the components of extrinsic curvature to $\Sigma$ corresponding to the interior and exterior metric can be obtained as:
\begin{align}
K^{-}_{\tau\tau}=-\phi_{r}(r,t)\frac{\sqrt{(1-\frac{b(r)}{r}}}{a(t)}\quad K^{-}_{\theta\theta}=ra(t)\sqrt{(1-\frac{b(r)}{r}}\quad K^{-}_{\psi\psi}=\sin^{2}\theta K^{-}_{\theta\theta}\\
K^{+}_{\tau\tau}=\frac{\frac{d^{2}v}{d\tau^{2}}}{\frac{dv}{d\tau}}-\left(\frac{dv}{d\tau}\right)\frac{M(v)}{R^{2}}\quad K^{+}_{\theta\theta}=R\left[\frac{dv}{d\tau}\left(1-\frac{2M}{R}\right)+\frac{dR}{d\tau}\right]\quad K^{+}_{\psi\psi}=\sin^{2}\theta K^{+}_{\theta\theta}
\end{align}
Equating $K^{-}_{\theta\theta}$ and $K^{+}_{\theta\theta}$ and using the matching conditions (\ref{bc1}), (\ref{bc2}) and (\ref{bc3}) one can obtain 
\begin{equation}
   M\overset{\Sigma}{=} \frac{ra(t)}{2}\left(\frac{b(r)}{r}+r^{2}e^{-2\phi}\dot{a}^{2}\right) \label{ms}   
 \end{equation}
which is the well know Misner-Sharp mass function, giving the total mass of the enclosed fluid within the hypersurface $\Sigma$ at any given time $t.$ It may be noted that the Misner-Sharp mass is also significant in the analysis of gravitational collapse to a black hole. Under the circumstance where the interior geometry collapses to a black hole, the outermost boundary of the interior metric (which is also the junction interface) becomes the black hole event horizon. When this happens $2M_{\Sigma}=R_{\Sigma},$ where $R_{\Sigma}=a(t)r_{b}$ is the areal radius of the co-moving outer boundary surface. If $t_{bh}$ is the time at which this happens, then for all $t<t_{bh},~2M_{\Sigma}<R_{\Sigma}.$ Hence (\ref{ms}) is important in determining the time of formation of ``Schwarzschild Singularity" of $2M_{\Sigma}=R_{\Sigma}.$  

Equating $K^{-}_{\tau\tau}$ and $K^{+}_{\tau\tau}$ and after some manipulation we obtain the time dependent differential equation 
\begin{equation}
\ddot{a(t)}a(t)-\phi_{t}\dot{a(t)}a(t)+\frac{\dot{a(t)}^{2}}{2}-\dot{a(t)}e^{\phi}\phi_{r}\sqrt{1-\frac{b(r)}{r}}+\frac{e^{2\phi}}{r^{2}}\left(\frac{b(r)}{2r}-\phi_{r}(r-b(r))\right)\overset{\Sigma}{=} 0.\label{me}
\end{equation}
It may be noted that using (\ref{fe2}) and (\ref{fe4}) equation (\ref{me}) can be written as 
\begin{equation}
p_{r}\overset{\Sigma}{=} q\frac{a(t)}{\sqrt{1-\frac{b(r)}{r}}}.
\end{equation}
Which shows that given a spherically symmetric distribution of radial heat dissipating fluid, the radial pressure at the junction must match the co-moving component of the heat quotient. The radial pressure will be zero only under the circumstance of $q=0$ \cite{san}.

\section{Solving the dynamical equation}

Since we are interested in understanding the collapsing dynamics of the above distribution of matter, we need to solve the time evolution equation (\ref{me}) which we rewrite as: 
\begin{equation}
\ddot{a}a+\frac{\dot{a}^{2}}{2}-\dot{a}\dot{\alpha}a-\dot{a}e^{\alpha}f_{\Sigma}(r)+e^{2\alpha}g_{\Sigma}(r)=0\label{fme}
\end{equation}
where $\phi(r,t)=\alpha(t)+\beta(r),~f_{\Sigma}(r)=e^{\beta(r)}\beta'(r)\sqrt{1-\frac{b(r)}{r}}$ and $g_{\Sigma}(r)=e^{2\beta(r)}\left[\frac{b(r)}{2r^{3}}-\frac{\beta'(r)(r-b(r))}{r^{2}}\right].$ Here we have separated the functions of $r$ and $t$ in $\phi(r,t)$ for ease of calculation. Further the functions $f_{\Sigma}(r)$ and $g_{\Sigma}(r)$ are evaluated for $r=r_{b}$ on the junction $\Sigma.$

\section{Solutions}

Equation (\ref{fme}) is a non-linear differential equation with unknown coefficients. Analytical solution of such equations are very difficult and not always possible. Further the equation contains unknown forms of the metric functions $\phi(r,t)$ and $b(r).$ As a result we have effectively three unknowns and one master equation. Evidently unique solution of all three unknowns require at least two more constraints. Hence we solve (\ref{fme}) subject to certain constraints as shown below:

\begin{itemize}

\item[{\bf S1}]: $a(t)=\frac{3}{2}(a_{0}t+a_{1})^{2/3}$

with $a_{0}$ and $a_{1}$ the integration constants. This solution is obtained by making $g_{\Sigma}(r)=0$ and $\dot{\alpha}a+e^{\alpha}f_{\Sigma}(r)=0.$

\item[{\bf S2}]: $a(t)=\frac{3}{2}\left(\frac{a_{0}}{\lambda}e^{\lambda t}+a_{1}\right)^{2/3}$
 
which is obtained by assuming $\dot{\alpha}=\lambda$, a constant parameter and $f_{\Sigma}(r)=g_{\Sigma}(r)=0,$ with $a_{0}$ and $a_{1}$ the integration constants.

\item[{\bf S3}]: $t=t_{0}+\frac{a}{2d}-\frac{a_{0}}{2d^{2}}a^{1/2}+\frac{a_{0}^{2}}{4d^{3}}\ln\left|a_{0}+2d\sqrt{a}\right|$

To obtain this solution we assumed $\dot{\alpha}=0=g_{\Sigma}(r).$ This reduces the (\ref{fme}) to:
\begin{equation}
\ddot{a}a+\frac{\dot{a}^{2}}{2}-d\dot{a}=0
\end{equation}
where $d=f_{\Sigma}e^{\alpha_{0}},~\alpha(t)=\alpha_{0},$ constant and $t_{0}$ being the integration constant. 

\item[{\bf S4}]: $a(t)=e^{\pm 2i wt}$

Here we have assumed $e^{\alpha(t)}=a(t)$ and  $f_{\Sigma}(r)=0$ with $w^{2}=\frac{g_{\Sigma}}{2}.$ 
\end{itemize}

Solutions S1-S4 are four different category of dynamical solutions that could be obtained for the equation (\ref{fme}) subject to the above constraints. Since the solution of (\ref{fme}) or in general (\ref{me}) depends on knowledge of the remaining three unknown functions $b(r),~\alpha(t)$ and $\beta(r),$ which makes the constraints in solutions S1-S4 essential.

\subsection{Solution {\bf S1}}

Here we discuss the various outcomes of solution S1 depending on the parameters of the model. We observe that in solution S1 the constant $a_{0}$ play the key role in determining the behaviour of $a(t).$ 

 \begin{figure}
\begin{minipage}[c]{0.4\linewidth}
\includegraphics[width=\linewidth]{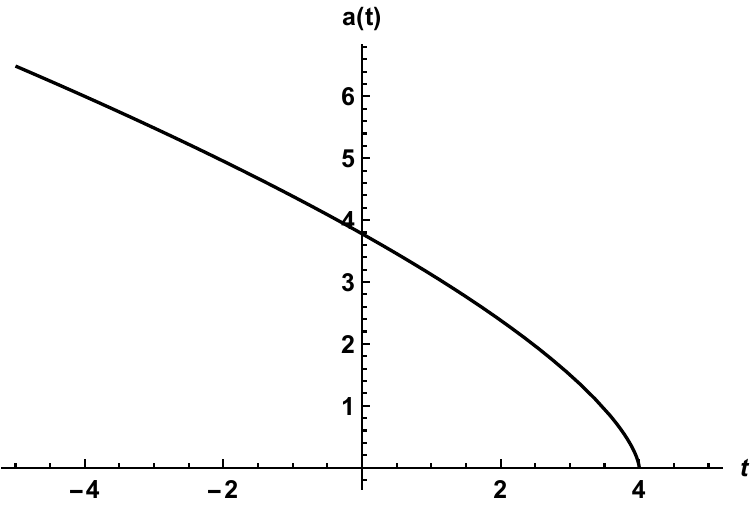}
\caption{The figure shows the values of $a(t)$ corresponding to $a_{0}<0$ and $a_{1}>0$ for solution S1.}
\end{minipage}
\hfill
\begin{minipage}[c]{0.4\linewidth}
\includegraphics[width=\linewidth]{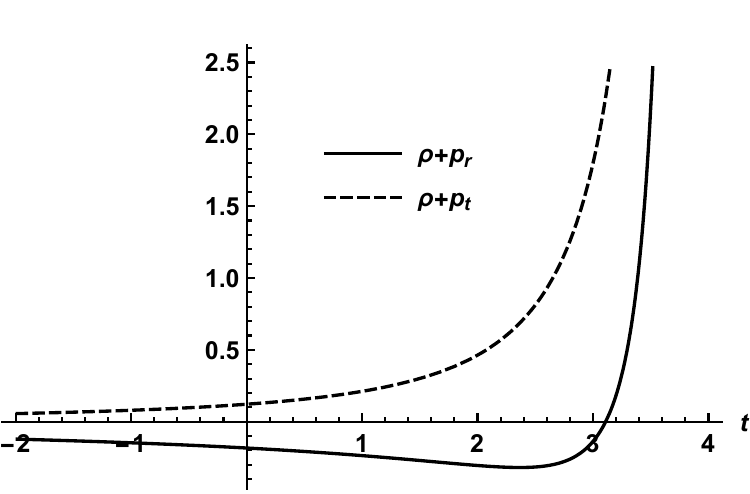}
\caption{The figure shows the energy conditions at the throat $r_{0}=0.7$ as $t$ changes corresponding to solution S1.}
\end{minipage}%
\end{figure}

\begin{figure}
\centering
\subfigure[]{\includegraphics[width=0.4\linewidth]{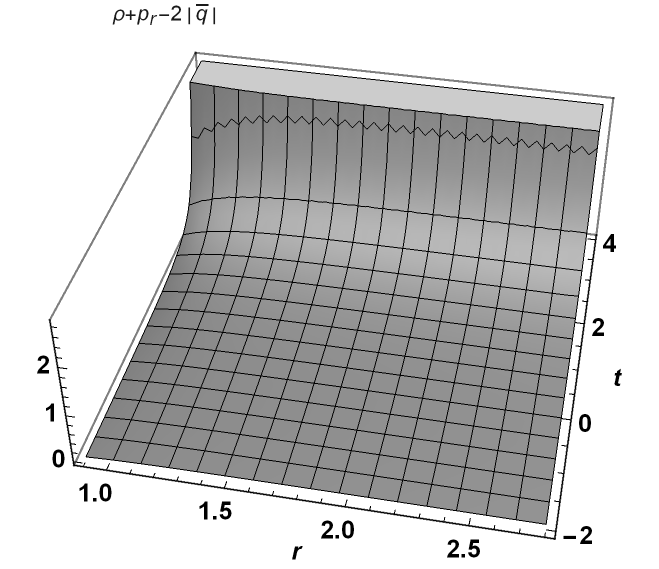}}
\subfigure[]{\includegraphics[width=0.4\linewidth]{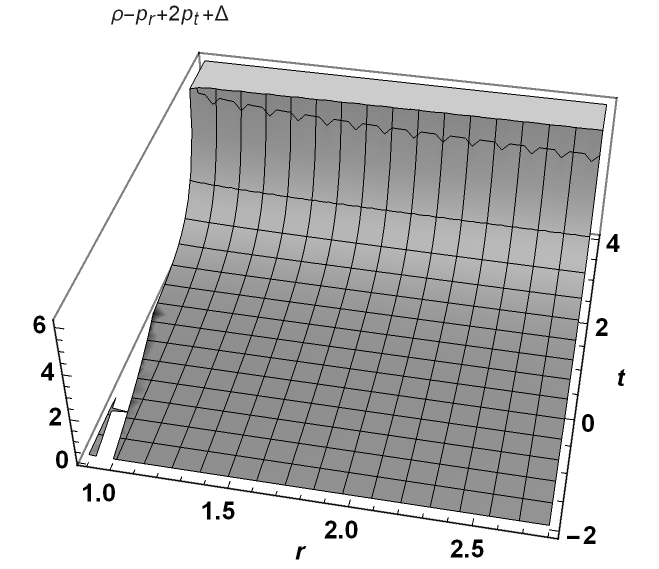}}
\caption{Here we show how the NEC as given by equations (24) and (25) change w.r.t $r$ and $t$ inside the interior metric for solution S1. Here $r_{0}<r<r_{b}$ which shows the variation of NEC in the interior metric at regions other than the throat $r_{0}$ and $q\neq 0.$}

\end{figure}

 \subsubsection{Collapsing Scenario}
 
\emph{{\bf The Scaling Function:}}  This is obtained for $a_{0}<0.$ In this case as $t\rightarrow-\infty,~a(t)\rightarrow\infty,$ while as $t\rightarrow 0,~a(t)\rightarrow\frac{3}{2}a_{1}^{\frac{2}{3}},~a_{1}>0$ and $a(t)\rightarrow 0$ at $t=-\frac{a_{1}}{a_{0}}(>0).$ This shows that for negative values of $a_{0}$ the interior distribution of matter will undergo rapid collapse to zero proper volume end state at $t_{s1}=-\frac{a_{1}}{a_{0}}.$ Figure 1 shows the evolution of the scale factor w.r.t time $t$ and for chosen value of parameters $a_{0}=-1$ and $a_{1}=4$ we see that the time of attaining zero proper volume is $t_{s1}=4.$

\emph{{\bf The Wormhole Vitals:}} The internal distribution of matter corresponds to a MT wormhole which is characterised by NEC violating matter near the throat $r_{0}.$ So we need to check the viability of solutions for the interior metric given for $r_{0}\leq r\leq r_{b}.$ The shape function describing the wormhole throat is assumed as $b(r)=\frac{r_{0}^{2}}{r}.$ With this throat description and using the constraints $g_{\Sigma}(r)=0$ and $\dot{\alpha}a+e^{\alpha}f_{\Sigma}(r)=0$ at $r=r_{b}$ we obtain the following results: \begin{align*}
\alpha(t)=&\ln\left[\frac{2f_{\Sigma}(r_{b})}{a_{0}}(a_{0}t+a_{1})^{\frac{1}{3}}+a_{2}\right]^{-1}\\
\beta(r)=&\frac{C}{r}\\
e^{2\phi(r,t)}=&e^{\frac{2C}{r}}\left\{\frac{2f_{\Sigma}(r_{b})}{a_{0}}(a_{0}t+a_{1})^{\frac{1}{3}}+a_{2}\right\}^{-2}
\end{align*} where $a_{2}>0$ is the integration constant and $C=-\frac{r_{0}^{2}r_{b}}{2(r_{b}^{2}-r_{0}^{2})}(<0).$ It may be noted that for the above configuration, the ensuing inhomogeneous and anisotropic fluid evolves to give a non-zero complexity factor $y_{tf}=\left(1-\left(\frac{r_{0}}{r}\right)^{2}\right)\left(\frac{3rC+C^{2})}{r^{4}a^{2}(t)}\right)-\frac{Cr_{0}^{2}}{a^{2}r^{5}},$ which might indicate an underlying instability of the system. Since the red-shift function $\phi(r,t)$ for a traversable MT wormhole is regular we obtain the following restriction: $t>t_{s1}-a_{0}^{2}\left(\frac{a_{2}}{2f_{\Sigma}(r_{b})}\right)^{3}$ which is the minimum $t$ and considered as initial time $t_{i}$.

\emph{\bf The Energy Conditions:} We are considering a dissipative distribution of fluid with dissipation characterized by the factor $q(r,t).$ The NEC for such a dissipative distribution of fluid is given as \cite{ec}:
\begin{align} 
\rho+p_{r}\geq 0\\
\rho+p_{r}-2|\bar{q}|\geq 0\label{nec1}\\
\rho-p_{r}+2p_{t}+\Delta\geq 0\label{nec2}\\
\Delta=\sqrt{(\rho+p_{r})^{2}-4\bar{q}^{2}}\\
\bar{q}=q\left(\frac{a(t)}{\sqrt{1-\frac{b(r)}{r}}}\right)\label{ec}
\end{align}
 Evidently without dissipation, the above conditions reduce to $\rho+p_{r}\geq 0$ and $\rho+p_{t}\geq 0.$ In the above considered case the dissipation coefficient is given by
 \begin{equation*}
 q(r,t)=\frac{16a_{0}Ce^{-\frac{C}{r}}}{27r^{2}}\left(\frac{\frac{2f_{\Sigma}(r_{b})}{a_{0}}(a_{0}t+a_{1})^{\frac{1}{3}}+a_{2}}{\left(a_{0}t+a_{1}\right)^{\frac{7}{3}}}\right)\left(1-\frac{r_{0}^{2}}{r^{2}}\right)
 \end{equation*}
  which is strictly positive for $r>r_{0}$ and $q(r_{0},t)=0.$ This means at the throat $r_{0}$ the NEC are given by $\rho+p_{r}\geq 0$ and $\rho+p_{t}\geq 0.$ Figure 2 shows the NEC values at the throat at different times. We see that as $t\rightarrow t_{s1}$ the NECs at the throat change from negative to positive. That is as volume of the sphere centred at the throat decreases, the matter stress tensors lining the throat becomes a NEC satisfying with $(\rho+p_{r},\rho+p_{t})\rightarrow+\infty$ as $t\rightarrow t_{s1}.$

However as we move away from the throat an analysis of the NEC show that the stress energy tensors lining the interior metric at all other $r_{0}<r\leq r_{b}$ is the NEC satisfying stress tensor with positive heat quotient. This means that the interior metric shrinks upon radial loss of heat energy from the shells surrounding the throat. In figure 3 we show the evolution of the NEC for $r\geq r_{0}.$ Thus, starting with a MT traversable wormhole, a scenario of gravitational collapse due to radially outgoing heat quotient is viable in regions away from the throat. But the existence of NEC violating matter complicates the apparent simple scenario. In order to explain the collapsing outcome and outgoing radial heat quotient we consider the several alternative explanations given for a dynamical wormhole throat \cite{vis,hay,rom}. In \cite{hay} a dynamic wormhole throat is considered as a temporal outer trapping horizon that can change into a spatial trapping horizon when the negative energy source of the wormhole throat fails and it starts to contract. Here we argue that the process of gravitational collapse initiates the process by which wormhole throat changes its temporal nature to get converted to a black hole. 

\begin{figure}
\begin{minipage}[c]{0.4\linewidth}
\includegraphics[width=\linewidth]{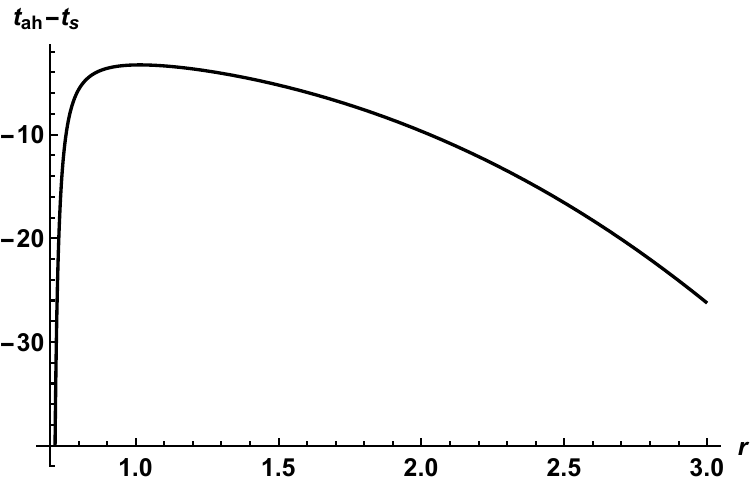}
\caption{The figure shows the change in equation (28) as $r$ varies from $r_{0}\leq r\leq r_{b}$ for the solution S1.}
\end{minipage}
\end{figure}
Here we list the various properties of the collapsing object:
\begin{itemize}
\item The time of formation of the apparent horizon $t_{ah1}$ provides the time of formation of trapping surfaces corresponding to subsequent outer shell collapse. The time to formation of apparent horizon is evaluated from $g^{\alpha\beta}\mathsf{R}_{,\alpha}\mathsf{R}_{,\beta}=0.$ For the given interior metric 
\begin{equation}
t_{ah1}-t_{s1}=-a_{2}^{3}a_{0}^{2}\left(e^{\frac{C}{r}}\sqrt{\frac{1}{r^{2}}-\frac{b(r)}{r^{3}}}+2f_{\Sigma}(r_{b})\right)^{-3}\label{ah1}
\end{equation}
As can be verified from figure 4, this is a monotonically decreasing function of negative values, showing that the apparent event horizon appearing before zero proper volume state for all the collapsing shells in $r_{0}<r\leq r_{b}.$ This is similar to the scenario depicted in the OSD models leading to a black hole. At the throat $r=r_{0}$ we see that the apparent horizon $t_{ah1}$ is formed at time $t_{i}.$ That is at the onset of collapse, as the negative energy source of the wormhole fails, the throat is converted to a spatial trapping horizon. As contraction ensues the subsequent shells collapse encasing the initial singularity at the throat by apparent horizons before total collapse at time $t_{s1}.$ 

\item Since the Misner-Sharp mass is used to locate the black-hole event horizon, we use the relation $2M_{\Sigma}=R_{\Sigma}$ at the outermost co-moving boundary $r_{b}$ to obtain the time $t_{bh1}$ of appearance of the ``Schwarzschild singularity" (outer trapping horizon). In fact it can be shown that the time of black hole formation is the same as the time at which apparent horizon is formed at $r_{b}.$ Thus from (\ref{ah1}), at $r=r_{b}$, we can see that $t_{i}<t_{bh1}<t_{s1}$ signalling the complete transformation of the wormhole into a black hole during the process of collapse. Further one can also observe that at $t=t_{i},~\frac{2M_{\Sigma}}{R_{\Sigma}|_{t_{i}}}=\left(\frac{r_{0}}{r_{b}}\right)^{2}<1$ clearly showing that initially the outer boundary was not a trapping surface.

\item The existence of a flux term is inevitable in our wormhole, due to the existence of non-zero tidal force terms. We can obtain the radially directed four acceleration as $F^{r}=\frac{\beta'(r)}{a^{2}}\left(1-\frac{b(r)}{r}\right).$ The direction of acceleration depends on the sign of the $\beta'(r),$ which is positive for S1. Hence we have $F^{r}>0.$ Correspondingly the energy flux $q(r,t),$ that depends on the radial four acceleration is positive, which in turn gives a scenario of collapse or $\dot{a}<0.$ Following \cite{rom}, we can say that the wormhole S1 is ``attractive" collapsing due to radial outflow of negative energy from the throat. This is synonymous to radial inflow of positive energy resulting in  instabilities that lead to collapse of the wormhole throat and formation of the black hole event horizon. Thermodynamically, the radial flux  will have negative temperature \cite{therm} due to phantom radiations. Thus using exact solution S1 we have outlined both numerically and analytically a scenario of gravitational collapse in a wormhole to a black hole. The figures 2-4 were obtained for $r_{0}=0.7$ and $r_{b}=1.75.$

\end{itemize}

\subsubsection{Expanding Scenario} Just as we obtained a contracting geometry for $a_{0}<0,$ we can get an expanding scenario with $a_{0}>0.$ With same solutions for $b(r),~\alpha(t),~\beta(r)$ and $a_{2}<0$ we again get an ``attractive" wormhole \cite{rom}, but this time with a negative flux $q(r,t)$ term, which in turn gives expansion. Thus here the wormhole inflates due incoming radial flux density. Here the wormhole simply inflates due to infusion of energy and then vanishes. For $a_{2}>0$ we get a highly dense distribution of matter that keeps expanding as $t\rightarrow\infty$ with corresponding stress tensors becoming zero at $\infty.$

\subsection{Solution {\bf S2}}

The various end states of dissipation for solution S2 will depend upon the nature of the constant parameters $a_{0}$ and $\lambda$ as elaborated below. 

 \subsubsection{Collapsing Scenario}
 
\emph{{\bf The Scaling Function:}} This is obtained for $\dot{a}<0,$ which is satisfied for $a_{0}<0$ and $\lambda<0,$ for real, physically viable solutions. Based on the values of $a_{1}$ we can have three different types of collapsing scenario. 
\begin{itemize}
\item {\bf C1: $a_{1}=0$} In this case S2 reduces to $a(t)=\frac{3}{2}\left(\frac{a_{0}}{\lambda}\right)^{\frac{2}{3}}e^{\frac{2}{3}\lambda t}.$ Here collapse happens between $-\infty<t<\infty$ with $a(t)\rightarrow 0$ as $t\rightarrow\infty.$
\item {\bf C2: $a_{1}>0$} Here collapse proceeds upto $a(t)=\frac{3}{2}a_{1}^{\frac{2}{3}}$ at $t\rightarrow\infty.$ 
\item {\bf C3: $a_{1}<0$} Here collapse proceeds upto some finite time $t_{s2}=\frac{1}{\lambda}\ln\left[-\frac{a_{1}\lambda}{a_{0}}\right]$ with $a(t_{s2})=0.$
\end{itemize}

\begin{figure}
\centering
\subfigure[]{\includegraphics[width=0.31\textwidth]{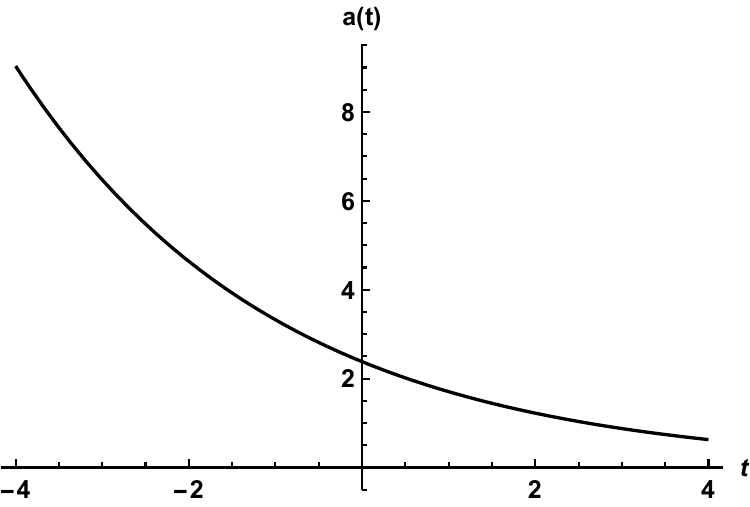}}
\subfigure[]{\includegraphics[width=0.31\textwidth]{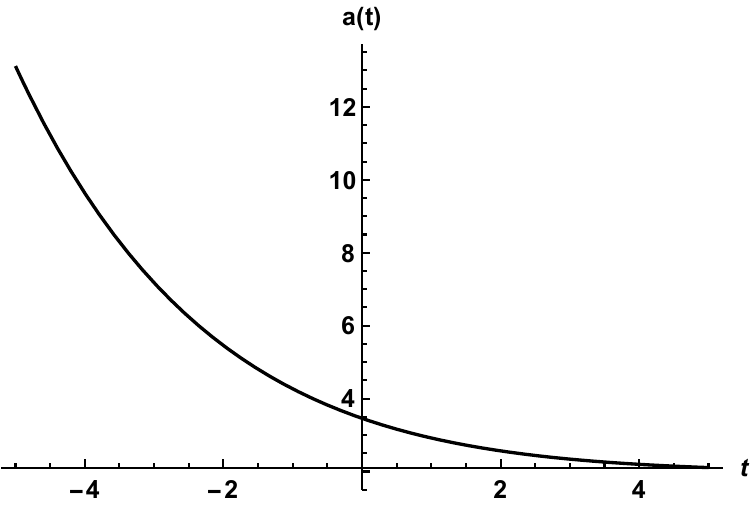}}
\subfigure[]{\includegraphics[width=0.31\textwidth]{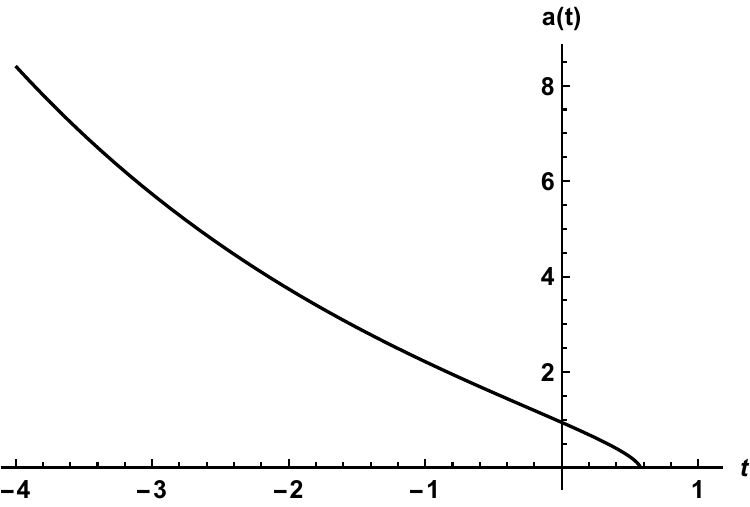}}
\caption{The figure shows the change in scale factor $a(t)$ for solution S2 and cases C1, C2, C3. Here subfigure (a) $a(t)$ in C1 and $a_{1}=0.$  (b) $a(t)$ in C2 and $a_{1}>0(=1.5).$  (c) $a(t)$ in C3 and $a_{1}<0(=-1.5).$ All these figures have been plotted for $a_{0}=-1,~\lambda =-0.5.$}
\end{figure}

\emph{\bf The Wormhole Description}: In order to obtain S2, we had assumed $\dot{\alpha}(t)=\lambda,$ which now gives $\alpha(t)=\lambda t.$ The constraints $f_{\Sigma}(r_{b})=0$ gives $\beta'(r_{b})=0,$ hence $q(r_{b},t)=0.$ From equation (\ref{fe4}) we further see that $q(r_{0},t)=0$ at the wormhole throat $r_{0}.$ Since $q(r,t)=0$ at both $r_{0}$ and $r_{b}$ we conclude that the radial flux density $q(r,t)=0$ for $r_{0}<r<r_{b}.$ This means $\beta'(r)=0$ or $\beta(r)=constant$ for all $r\in[r_{0},r_{b}].$ Thus $\phi(r,t)$ is independent of $r$ and hence 
\begin{equation*}
e^{2\phi(r,t)}=e^{2\lambda t}
\end{equation*}
That $\phi_{r}(r,t)=0$ will have important consequence from the stability perspective of this MT wormhole. As already stated, under homologous conditions $\phi_{r}=0$ gives geodesic flow with $y_{tf}=0,$ which is what we obtain here. Thus the corresponding fluid gives a ``simple" wormhole which is stable in the presence of inhomogeneity and pressure anisotropy. The third constraint $g_{\Sigma}(r_{b})=0$ give $b(r_{b})=0.$  Satisfying the wormhole throat criterion together with $b(r_{b})=0$ give 
 \begin{align*}
 b(r)=rb_{0}\left[\left(\frac{r_{b}}{r}\right)^{2}-1\right]\\
 b_{0}=\frac{r_{0}^{2}}{r_{b}^{2}-r_{0}^{2}}>0
 \end{align*} With these solutions we study the ultimate fate of the wormhole for the scenarios C1, C2 and C3.
  
 \vspace{1em}

\emph{\bf C1:} This is a classic case of contracting wormhole with zero proper volume as $t\rightarrow\infty.$ The NEC violation of the stress energy tensors of the wormhole changes to one that is NEC satisfying with $\rho,p_{r},p_{t}\rightarrow\infty$ as $t\rightarrow\infty.$ With the above solution of $a(t)$ and $e^{\phi(r,t)}$ we can re-write the interior metric in term of the conformal time $d\tau=a(t)dt$ as 
\begin{equation*}
dS_{-}^{2}=-d\tau^{2}+A^{2}(\tau)\left(\frac{dr^{2}}{1-b(r)/r}+r^{2}(d\theta^{2}+\sin^{2}\theta d\psi^{2})\right)
\end{equation*}
 with $A(\tau)=\frac{3}{2}(a_{0}\tau)^{\frac{2}{3}}.$ Thus the interior metric is conformal to the inhomogeneous matter dominated FRW universe, (here the conformal time $\tau\in(-\infty,0)$ is the past time because we are looking at a collapsing scenario with negative $\lambda$). In the outermost shell/junction at $r=r_{b}$ this is essentially the flat matter dominated FRW metric. Here the collapsing object represents a decelerated shrinking universe, that finally collapses to zero volume as $\tau\rightarrow 0.$
 
 \vspace{1em}

\emph{\bf C2:} In this case the wormhole contracts and becomes a static geometry with finite proper volume. This is a case of evolving wormhole collapsing to a static wormhole geometry, has been already obtained in \cite{sksc}. On inspecting the energy conditions at the wormhole throat we observe that the throat matter can satisfy or violate the NEC based on the choice of the parameters $r_{0},r_{b}$ and $a_{1}.$ 

 \begin{figure}
\begin{minipage}[c]{0.4\linewidth}
\includegraphics[width=\linewidth]{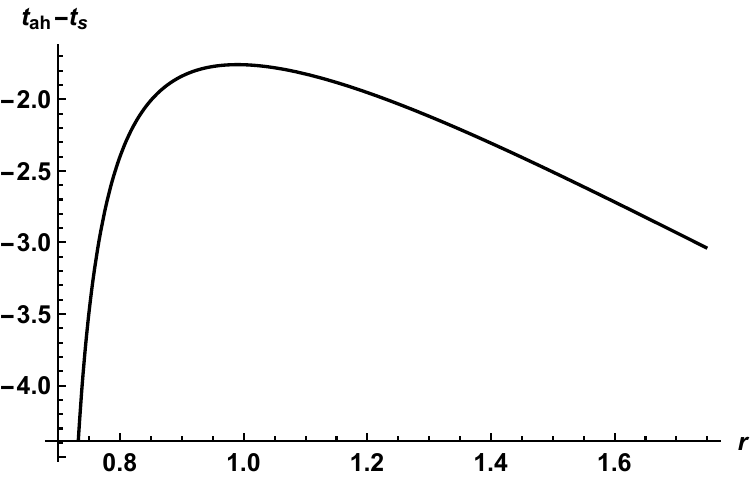}
\caption{The figure shows the values of $t_{ah2}-t_{s2}$ corresponding to $a_{0}<0,~a_{1}<0$ for solution S2 and case C3. Here $t_{s2}$ is evaluated to 0.58 and $r_{0}=0.7$ with $\lambda=-0.5$ }
\end{minipage}
\hfill
\begin{minipage}[c]{0.4\linewidth}
\includegraphics[width=\linewidth]{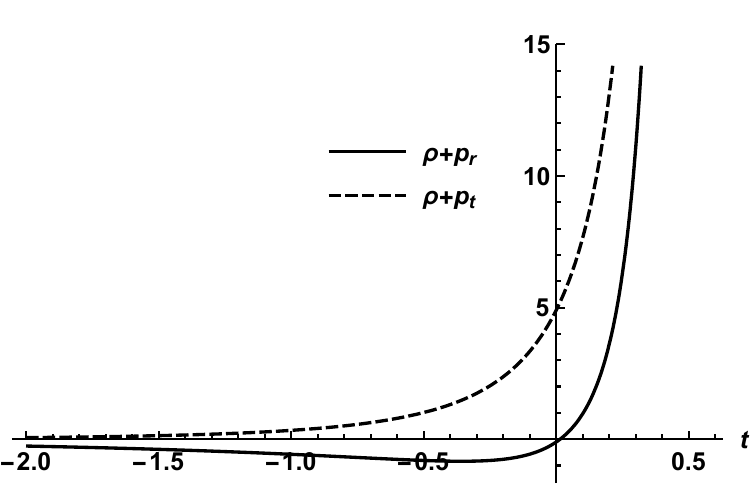}
\caption{The figure shows the energy conditions at the throat $r_{0}=0.7$ as $t$ changes corresponding to solution S2. Here also $t_{s2}=0.58$ and $\lambda=-0.5.$}
\end{minipage}%
\end{figure}

\emph{\bf C3:} Here collapse ends with zero proper volume in some finite time $t_{s2}.$ The evolution of the scale factor $a(t)$ for this case has been shown in figure 5(c). Here we discuss some of the features of the collapsing geometry:
\begin{itemize}
\item The apparent horizon for subsequent collapsing shells is given by: 
\begin{equation}
t_{ah2}=\frac{1}{\lambda}\log\left[- \lambda a_{0}^{2}r^{3}\left(1-\frac{b(r)}{r}\right)^{-\frac{3}{2}}-\frac{a_{1}\lambda}{a_{0}}\right]\label{ah2}.
\end{equation}
 Again $t_{ah2}-t_{s2}$ is a monotonically decreasing function of negative values, showing that the consecutive shells upon collapse are always hidden under the apparent horizon as demonstrated in figure 6. As in S1, at $r=r_{0},~t_{ah2}=-\infty$ which is the time of initiation of collapse. This is usual considering that as collapse begins the trapped surface at $r_{0}$ changes its temporal nature to spatial and is always hidden under the apparent horizon. It is known that wormholes in general are highly unstable structures, hence there can be several reasons facilitating its collapse, like Gaussian pulses \cite{hay2}, spontaneous creation/annihilation of matter particles or entangled particles \cite{lob} or quantum fluctuations. In the absence of radial dissipative processes and stability of matter tensors we conclude that collapse is realized due to the inherent instability of the wormhole throat.
 
\item As we have shown in S1, here also we observe that the time of black hole formation is same as the time when apparent horizon is formed at the outermost boundary surface $r_{b}.$ Thus using (\ref{ah2}) we can see that the time of black hole formation $t_{i}<t_{bh2}<t_{s2}$ where $2M_{\Sigma}=R_{\Sigma}$ signalling the formation of trapping horizon or the event horizon corresponding to the collapse and black hole formation. At $t=t_{i},~\frac{2M_{\Sigma}}{R_{\Sigma}}\rightarrow 0(<1),$ showing the regular initial configuration of $r_{b}.$ 
 
\item It is important to understand the evolution of the energy conditions of the matter stress tensors in the interior metric. The NEC at the throat $r=r_{0}$ show that as $t\rightarrow-\infty,$ that is, at the initiation of collapse, the throat is lined with NEC violating matter but only in finitely small quantity. As $t\rightarrow t_{s2}$ this is soon changed into a NEC satisfying matter tensor. This is demonstrated in figure 7. For any other $r_{0}<r\leq r_{b}$ the evolution of the energy density and the NEC follows same pattern with marginal negative value for $t<0,$ which changes to positive matter with collapse. Since $q=0$ for all $r_{0}<r\leq r_{b},$ the NEC will be given by $\rho+p_{r}\geq 0$ and $\rho+p_{t}\geq 0$ instead of the equations (\ref{nec1}) and (\ref{nec2}). The figure 8 shows the evolution of the NEC at different shells starting from $r=r_{0}$ and upto $r=r_{b}.$ (All figures for C3 has been plotted for $a_{0}=-1,~a_{1}=-1.5,~\lambda=0.5,~r_{0}=0.7,~r_{b}=1.75$).

\end{itemize}
 
Solution S2 is very interesting in the sense that, we started with a stable configuration, and probably as a corresponding consequence we could obtain a couple of singularity free configurations apart from the black hole singularity. Contrast of scenarios C1 and C2 with C3 reiterates that instability within a system can result in the formation of black holes.

\begin{figure}
\centering
\subfigure[]{\includegraphics[width=0.4\textwidth]{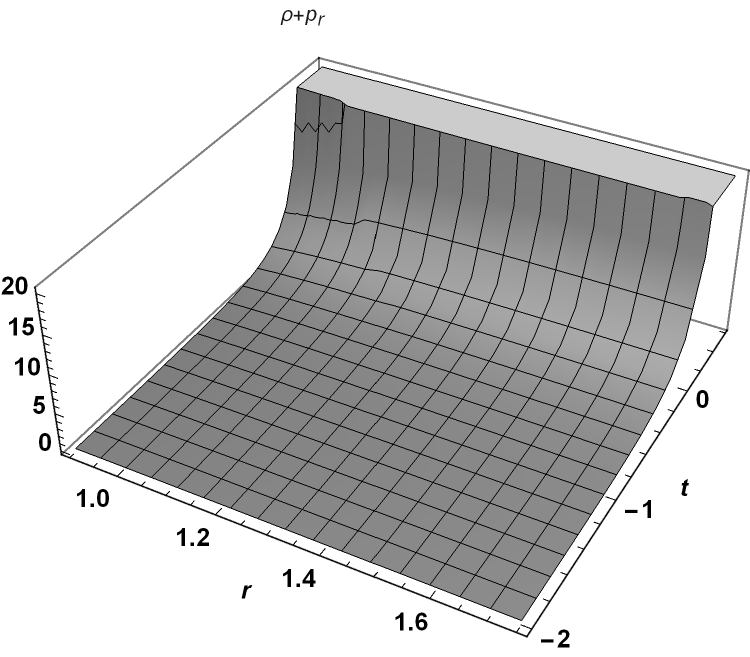}} 
\subfigure[]{\includegraphics[width=0.4\textwidth]{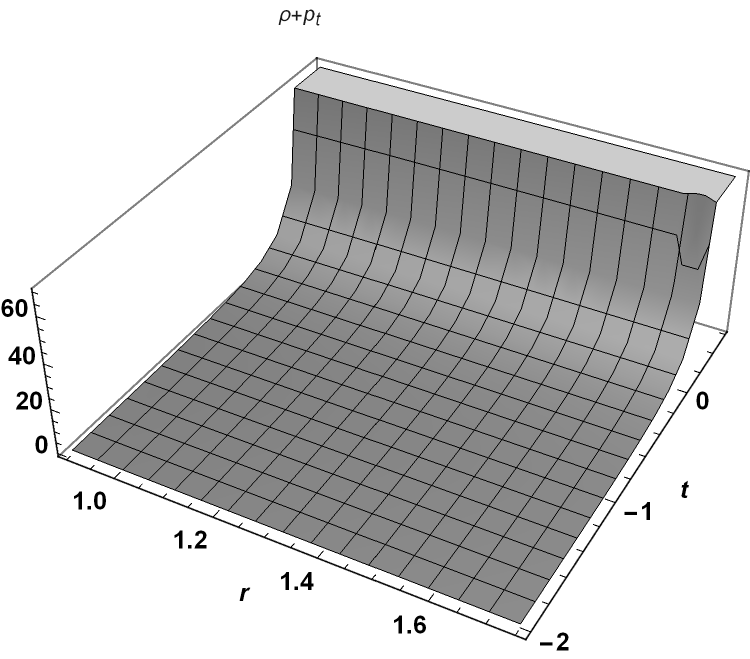}}
\caption{Here we show how the NEC  wrt $r$ and $t$ inside the interior metric for solution S2 and case C3. Here $r_{0}<r<r_{b}$ which shows the variation of NEC in the interior metric at regions other than the throat $r_{0}$ and $q= 0.$}
\end{figure}

 \subsubsection{Expanding Scenario}
 
  All inflating geometries are obtained for $a_{0}>0$ and $\lambda>0.$ For $a_{1}=0$ we get an inflating wormhole with a De-Sitter scale factor $a(t)=\frac{3}{2}\left(\frac{a_{0}}{\lambda}e^{\lambda t}\right)^{\frac{2}{3}}$ \cite{rom}. For $a_{1}>0$ expansion proceeds upto $a\rightarrow\infty$ for $t\rightarrow\infty.$ For $a_{1}<0$ expansion is initiated only from time $t=\frac{1}{\lambda}\log\left(\frac{-a_{1}\lambda}{a_{0}}\right)$ such that $-a_{1}\lambda<a_{0}.$ At this point $a(t)=0$ and expansion proceeds upto $t\rightarrow+\infty$ with $a(t)\rightarrow+\infty.$

\subsection{Solution {\bf S3}}

\subsubsection{Collapsing Scenario}

\emph{{\bf The Scaling Function:}} In S3, the scale factor $a(t)$ is obtained implicitly in terms of $t,$ as \begin{equation}
 t=t_{0}+\frac{a}{2d}-\frac{a_{0}}{2d^{2}}a^{1/2}+\frac{a_{0}^{2}}{4d^{3}}\ln\left|a_{0}+2d\sqrt{a}\right|\label{a3}
 \end{equation}
  where $d=f_{\Sigma}e^{\alpha_{0}},~\alpha(t)=\alpha_{0},$ constant and $t_{0}$ being the integration constant. From $\dot{a}(t)=a_{0}a^{-\frac{1}{2}}+2d$ we get $\dot{a}=0$ for $a(t)=\left(\frac{a_{0}}{2d}\right)^{2}$ which is a constant and occurs at $t\rightarrow-\infty.$ Further we find that at $a(t)=0,~t=t_{0}+\frac{a_{0}^{2}}{4d^{3}}\log|a_{0}|=t_{s3}.$ This shows that $a(t)$ decreases to zero as $t$ increases from $-\infty$ to $t_{s3}.$ Thus a collapsing geometry will be recognized for viable values of the parameter $a_{0}<0$ and $d>0$ with collapse proceeding to zero proper volume in finite time $t_{s3}.$ The change in $t(a)$ wrt $a$ is shown in figure 9.

\emph{\bf The Wormhole Description:} We found the above solution using the conditions $\dot{\alpha}(t)=0$ which gives $\alpha(t)=\alpha_{0}$ a constant. Further we observe that taking $b(r)$ and $\beta(r)$ same as solution S1 satisfies the condition $g_{\Sigma}(r_{b})=0.$ This gives $e^{2\phi(r,t)}=e^{2\alpha_{0}}e^{\frac{2C}{r}},~ C$ defined as in S1. Again this means the existence of a possible inherent instability in the fluid distribution. Since $\beta'(r)\neq 0$ in this case we can expect a radially directed dissipative heat quotient. Here
\begin{equation*}
q(r,t)=\frac{2\dot{a}}{a^{3}}e^{-\alpha_{0}}e^{-\frac{C}{2r}}\left(\frac{C}{r^{2}}\right)\left[1-\left(\frac{r_{0}}{r}\right)^{2}\right]>0
\end{equation*}
 for $\dot{a}<0.$ Hence collapsing scenario is aided by an outward directed heat quotient. At $r=r_{0},~q(r,t)=0.$ 

Below we comment on the nature of the end singularity:
\begin{itemize}
\item Using $g^{\alpha\beta}\mathsf{R}_{,\alpha}\mathsf{R}_{,\beta}=0$ we get the value of the scale factor at the time of formation of apparent horizon corresponding to different shells $r_{0}\leq r\leq r_{b}$ as 
\begin{equation}
a(t_{ah3};r)=a_{0}^{2}\left[2d+\frac{e^{\alpha_{0}}e^{\frac{C}{r}}}{r}\sqrt{1-\left(\frac{r_{0}}{r}\right)^{2}} \right]^{-2}\label{ah3}.
\end{equation}
Using this in (\ref{a3}) we get: 
 \begin{equation}
 t_{ah3}-t_{s3}=\frac{a(t_{ah3};r)}{2d}-\frac{a_{0}}{2d^{2}}\sqrt{a(t_{ah3};r)}+\frac{a_{0}^{2}}{4d^{3}}\ln |1+\frac{2d}{a_{0}}\sqrt{a(t_{ah3};r)}|\label{tah3}.
 \end{equation} 
 From figure 10, for given values of the parameters $a_{0},~\alpha_{0}$ and $r_{b}$ we get a monotonically decreasing negative valued function in regions away from the wormhole throat. This shows that the singularity resulting from collapse is always covered by the apparent horizon.
 
\item  As before, the time of appearance of Schwarzchild singularity at the outer co-moving boundary at $r=r_{b}$ coincides with the time of apparent horizon formation at $r_{b}$. From (\ref{ah3}) we can obtain the value of $a(t_{ah3};r_{b})=a_{bh}.$ Then from equation (\ref{tah3}) we can obtain $t_{bh3}=t_{ah3}|_{r_{b}}$ and verify that $t_{i}<t_{bh3}<t_{s3}.$ Thus as collapse is initiated, the wormhole throat changes to the spatial trapping horizon (which is covered by the apparent horizon) converting to a black hole, with subsequent shells collapsing from the boundary towards the throat. 

\item At the throat and at $t\rightarrow-\infty,$ the NECs are given as 
 \begin{align*}
 \rho+p_{r}=&-\frac{32d^{4}}{a_{0}^{4}r_{0}^{2}}<0\\
 \rho+p_{t}=&-\frac{16Cd^{4}}{r_{0}^{3}a_{0}^{4}}>0.
 \end{align*}
But as $t\rightarrow t_{s3},$ or, as $a\rightarrow 0$ both the NECs diverge. For the other intermediate shells we can have NEC violation or NEC might hold depending upon the parameters $a_{0},~\alpha_{0}$ and $r_{b}.$
\end{itemize}
 
\begin{figure}
\begin{minipage}[c]{0.4\linewidth}
\includegraphics[width=\linewidth]{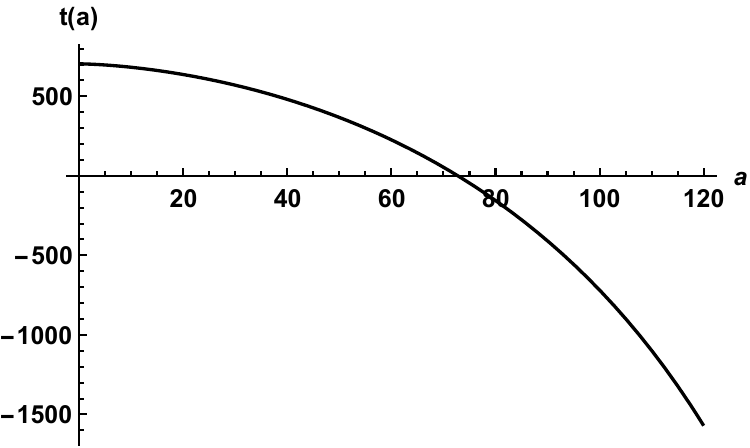}
\caption{The figure shows the values of $t(a).$ We have taken $a_{0}=-1.2$ and $d$ evaluated as 0.045 for $\alpha_{0}=0,~r_{0}=0.7,~r_{b}=1.75.$ For these parameters the time at which $a=0$ is given by $t_{s3}=703.86$ }
\end{minipage}
\hfill
\begin{minipage}[c]{0.4\linewidth}
\includegraphics[width=\linewidth]{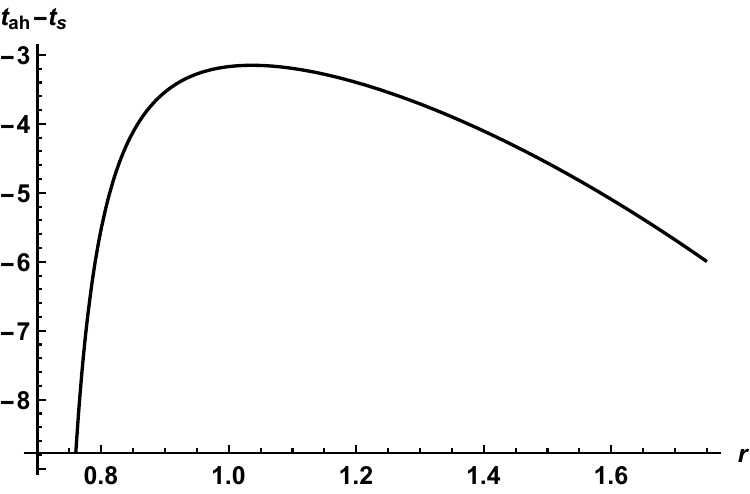}
\caption{The figure shows the values of $t_{ah3}-t_{s3}$ corresponding to $a_{0}=-1.2$ and $d$ evaluated as 0.45 for $\alpha_{0}=0,~r_{0}=0.7,~r_{b}=1.75.$}
\end{minipage}%
\end{figure}
 
We have once again provided a scenario of collapsing wormhole solution. Wormholes are highly unstable structures due to the requirement of negative energy at the throat, hence they are prone to collapse. We observe that the wormhole throat degenerates as soon as instabilities are produced due to collapse. In this example we have a radial outward dissipative term, which can contribute to the production of instabilities resulting into its collapse. It can be noted that solutions similar to S3 have been discussed in previous cases of radial dissipative heat collapse for anisotropic stars \cite{bonn}. This result was replicated in various other works describing collapse in anisotropic and radially dissipative distribution of matter (for details see \cite{s3} and references therein). This shows that the S3 is a generic collapse solution for any spherically symmetric anisotropic distribution of matter with radially dissipative flux. We further state that such a solution can also exhibit expansion depending on physical processes bringing the change of the underlying fluid.

\begin{figure}
\centering
\subfigure[]{\includegraphics[width=0.3\textwidth]{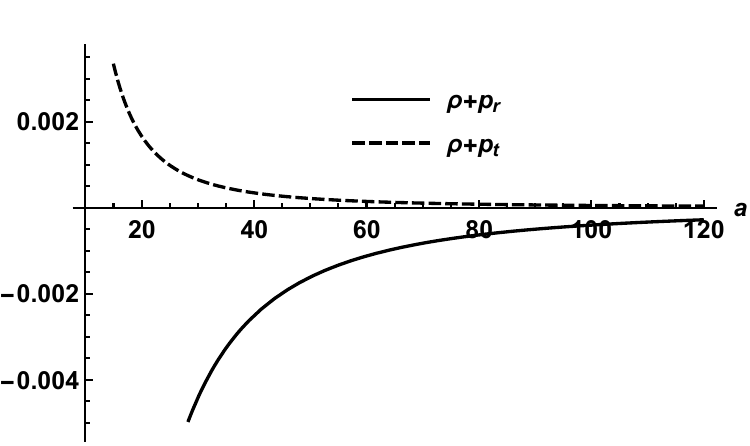}}
\subfigure[]{\includegraphics[width=0.31\textwidth]{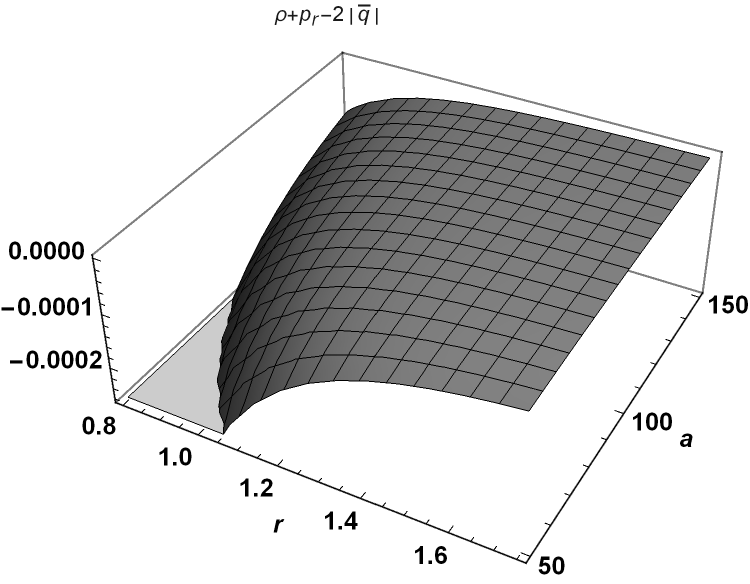}}
\subfigure[]{\includegraphics[width=0.31\textwidth]{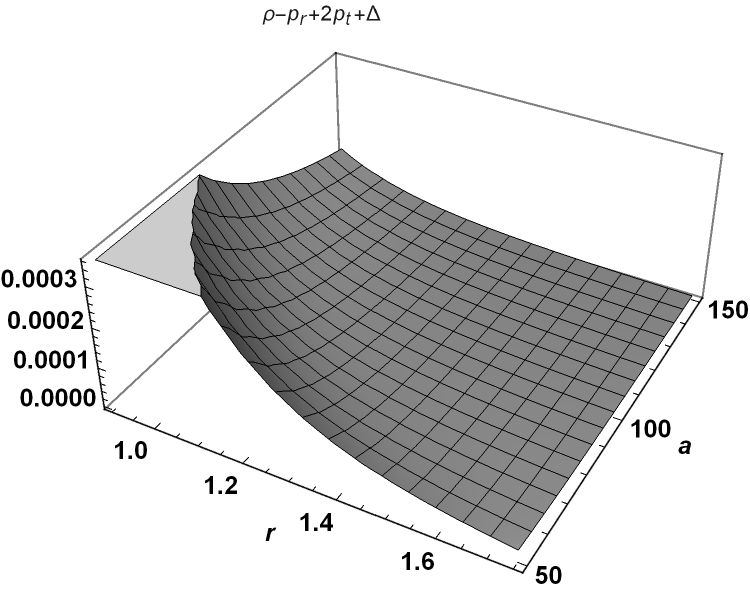}}
\caption{(a) This shows the NEC at the wormhole throat of solution S3 with parameter values as listed in figures 9 and 11, (b) and (c) These give the NEC given by equations (24) and (25) in the interior metric for solution S3 and for $r_{0}<r\leq r_{b}.$}
\end{figure}

\subsubsection{Expanding Scenario} 
Just as collapse geometry was obtained for $\dot{a}\leq 0$ one can also obtain expanding scenario for $\dot{a}\geq 0.$ This solution however gives parameter independent expansion. At $t\rightarrow-\infty$ we start with a static wormhole configuration and $q=0.$ As time increases, it is entirely the physical processes that determine the direction of flow of $q$ or whether $\dot{a}<0/(>0).$ For negative dissipative radial heat flow get an expanding geometry instead of collapse. So the Solution S3 differs from the previous two solutions, in the sense that, here we can get either contraction or expansion depending upon the kind of physical processes determining the direction of radial dissipation. In S1 or S2 we had the mathematical switches in the form of parameters which could give us our desired solutions based on our choices.

\subsection{Solution S4}

Solution S4 is different from all the solutions that we have discussed so far. We can get $a(t)=\sin^{2}\omega t$ or $a(t)=\cos^{2}\omega t$ depending on the initial conditions. This clearly shows that $\dot{a}(t)$ will be an oscillating function, which gives an oscillating contracting/expanding wormhole. The red-shift function $e^{2\phi(r,t)}=a^{2}$ and the shape function can be found as $b(r)=\frac{r_{0}^{2}}{r}.$ Thus beginning with a collapsing wormhole configuration, we get a black hole or an initial black hole changing into a wormhole. This is a cyclical ever existing wormhole-black hole pair. Since $\beta'(r)=0$ the corresponding radially dissipative heat flux is absent in this case and that the system being homologous and geodesic, will be stable in the presence of anisotropy and inhomogeneity.

\begin{figure}
\centering
\subfigure[]{\includegraphics[width=0.44\textwidth]{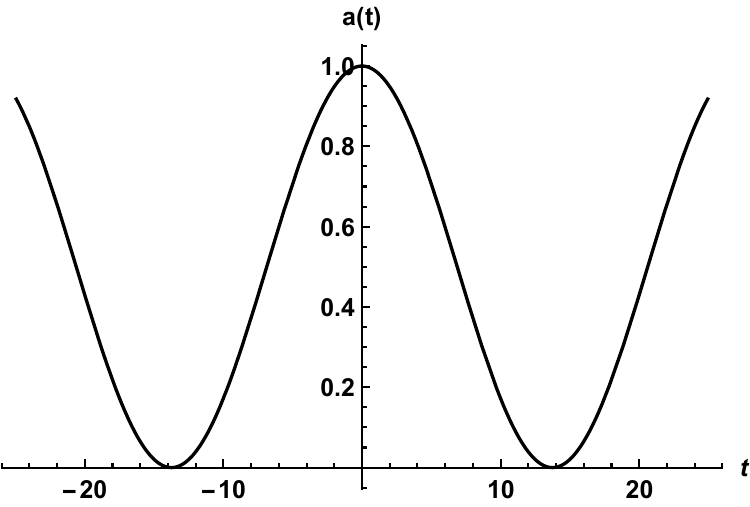}}
\subfigure[]{\includegraphics[width=0.44\textwidth]{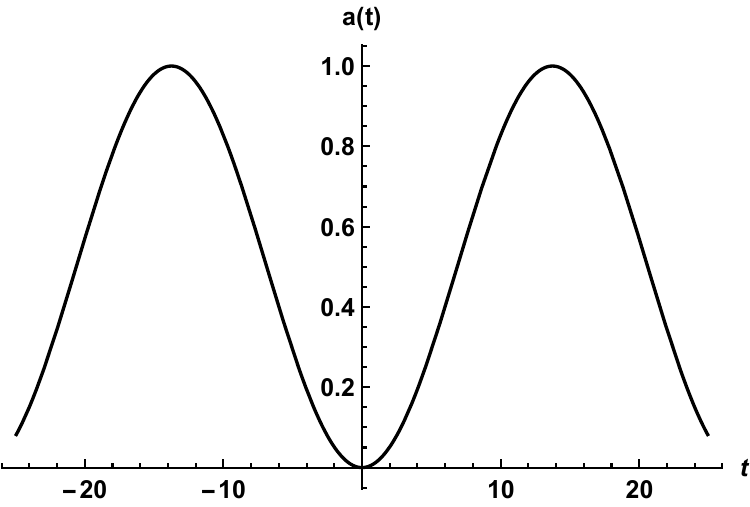}}
\subfigure[]{\includegraphics[width=0.44\textwidth]{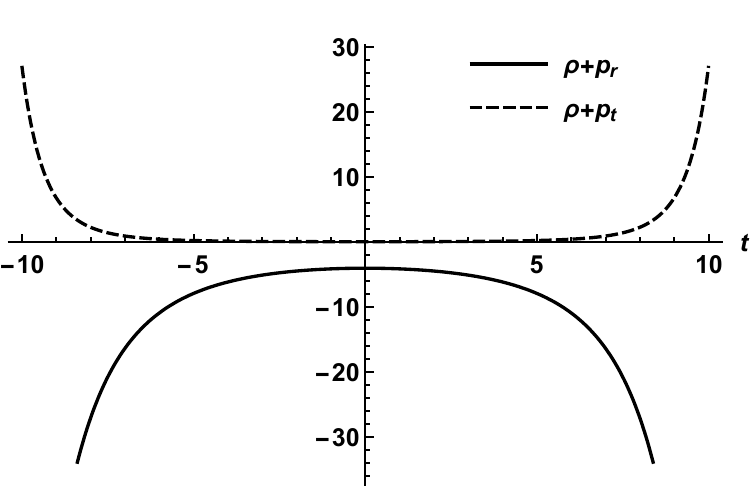}}
\subfigure[]{\includegraphics[width=0.44\textwidth]{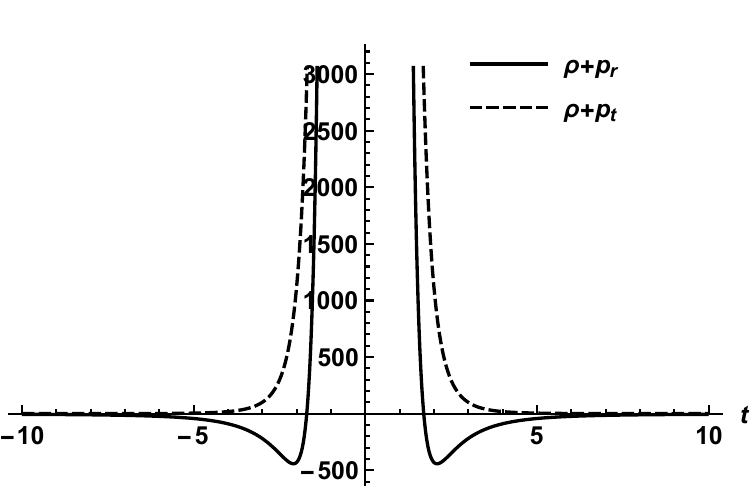}}

\caption{(a) This shows $a(t)=\cos^{2}\omega t.$ (b) The evolution of $a(t)=\sin^{2}\omega t$ (c) The NEC at the throat $r_{0}=0.7$ for $a(t)=\cos^{2}\omega t$ (d) The NEC at $r_{0}=0.7$ for $a(t)=\sin^{2}\omega t$}
\end{figure}

One can see that a wormhole structure is obtained at $a(t)=1$ with finite red-shift factor or no event horizon. While for $a(t)=0,$ the red-shift factor is infinite emphasising that the wormhole has transitioned into a black hole with a event horizon. Such black hole-wormhole pair have been discussed previously in the literature. In fact in \cite{nsk1} static black hole configurations developed in the galactic core were identified as wormhole entrances. In the presence of an electromagnetic field, the authors of \cite{nsk1} show how wormhole can manifest as a black hole. Thus the above solution can give a mathematical representation of such astrophysical wormhole, which periodically manifests itself as a black hole. In the broader context, one might also consider the work \cite{nsk2} where a singularity free, oscillating cosmological model of an ever-existing cyclical universe was developed for $a(t)=e^{\pm it}.$ Although in the light of recent observational data such a cosmological scenario might not be viable, yet one can optimistically comment on the probable existence of such periodically interchanging singularities in the universe.

Since we are claiming the formation of a black hole and back again to a wormhole, as before we can compute the time of black hole formation, that is the time the outer horizon $r_{b}$ becomes an event horizon. For $a(t)=\sin^{2}\omega t$ we obtain the time of black hole formation as $t_{bh4}=\frac{1}{\omega}\cot^{-1}\left[-\frac{1}{2\omega r_{b}}\left(1-\left(\frac{r_{0}}{r_{b}}\right)^{2}\right)\right]$ (here $\cot\omega t<0$ in $-\frac{\pi}{2\omega}<t<0$ which is the contracting phase with $\dot{a}<0$). Clearly $t_{bh4}<t_{s4}(=0).$ For $0<t<\frac{\pi}{2\omega}$ we get expanding configuration with $\dot{a}>0$ and in that case we again get the wormhole and at the outermost boundary $r_{b}, ~\frac{2M_{\Sigma}}{R_{\Sigma}}<1.$ This clearly shows the continuous interconversion of the nature of the boundary at $r_{b}$ and hence that of the ensuing configuration. (Similar calculations can be shown for $a(t)=\cos^{2}\omega t.$) 

From figure 12a we observe that $a(t)=\cos^{2}\omega t$ initially gives wormhole structure and then collapses into a black hole. This can be further sene from figure 12c. Similarly we see from figure 12b, that  $a(t)=\sin^{2}\omega t$ initially is a black hole that changes into a wormhole as can be further corroborated using figure 12d.

\section{Conclusion}

Our aim in this paper was to study the possibilities of collapse in the inhomogeneous MT metric. The static MT metric is generally known for harbouring a traversable wormhole. Tailoring the wormhole for the purpose of traversal necessitated a geometric manipulation rendering the matter-stress tensors singular at its throat. However the dynamic counterpart was altogether different and has greater mathematical and physical flexibility of interpretation at the throat. Harbouring on this we considered the radially dissipative evolving MT wormhole and matched it with the Vaidya metric to obtain proper dynamical equation. The corresponding matter distribution was assumed as inhomogeneous, anisotropic, dissipative and shear free. The presence of anisotropy with dissipation and shear free conditions made the configuration prone to instability. This made a discussion on stability imperative, which we did using that recently proposed complexity factor  for compact spherically symmetric configuration of matter. Using \cite{herr8} we could show that solutions S2 and S4 were ``simple" and stable, while no definitive conclusion on stability could be made for S1 and S3 based on their ``complexity".

The dynamical equation obtained by matching junction conditions involved unknown  functions for the wormhole throat and tidal forces. Hence solving such an equation required setting up of constraints which lead to different solutions. Interestingly this lead to several collapsing outcomes. Borrowing from the definition in \cite{hay} we developed the mathematics of wormhole changing to a black hole and vice-versa.  S1 and S3 collapsing solutions were accompanied by outward directed radial flux term, that lead to the change. While in S2 collapsing scenarios the radial flux term was absent. Thus our solutions could depict two scenarios of collapse mediated by different effects. In S2-C1 we obtained a collapsing scenario which could be interpreted as the collapsing past conformal matter-dominated inhomogeneous universe, a scenario similar to the familiar cosmological big crunch. In S2-C2 our solutions show the conversion of a evolving wormhole into a static wormhole. A result which was recently discussed in \cite{sksc}. However the interpretation of the collapse was modified in \cite{sksc} due to non-zero radial flux. Solution S4 gave mathematical representation of the  interchanging wormhole-black hole pair. 

We have provided a comprehensive mathematical description of the several probable background dynamics that could lead to a black hole formation from a wormhole. Although black hole has been detected observationally, wormholes are elusive. As our results point, one possible reason could be that they decay into a black hole almost at the instant they are formed. Further our solutions provided simple mathematical description of the background mechanism of how wormhole throat could change into a black hole. The results also provide a comparative description of different dynamics that can lead to collapse. Solution S3 has been previously related to gravitational collapse in different spherically symmetric scenarios was also obtained as a solution here. This shows that S3 is a generic solution of the dynamical equation resulting from junction matching criterion. 

Our solutions S1-S3 also can describe an inflating wormhole. This would depend on the parameters present in the solution. However inflating wormhole is unlikely as, that would mean conversion of normal matter to exotic matter. Nonetheless mathematically relevant expanding geometries could be built from our solutions. 

\section{ Acknowledgements:}

SN acknowledges UGC, Government of India, for financial assistance through junior research fellowship (NTA ref.no. 231610097492). SB acknowledges IUCAA, Pune, India, where the project was initiated.

\end{document}